\documentclass{aa}
\usepackage{txfonts}
\usepackage{graphicx}
\usepackage{natbib}

\begin{document}

\title{High time resolution observations of solar H$\alpha$ flares - II\\
 Search for signatures of electron beam heating}

\author{K. Radziszewski\inst{1} \and P. Rudawy\inst{1} \and K. J. H. Phillips\inst{2}}

\institute{ Astronomical Institute of Wroc{\l}aw University, 51-622
Wroc{\l}aw, ul. Kopernika 11, Poland\\
\email{radziszewski@astro.uni.wroc.pl; rudawy@astro.uni.wroc.pl}
 \and Mullard Space Science Laboratory, Holmbury St Mary, Dorking, Surrey RH5 6NT, United Kingdom\\
\email{kjhp@mssl.ucl.ac.uk} }

\offprints{K. Radziszewski \email{radziszewski@astro.uni.wroc.pl}}

\date{Accepted for publication - 13 September 2011   }

\abstract{} {The H$\alpha$ emission of solar flare kernels and
associated hard X-ray (HXR) emission often show similar time
variations but their light curves are shifted in time by energy
transfer mechanisms. We searched for fast radiative response of the
chromosphere in the H$\alpha$ line as a signature of electron beam
heating.} {We investigate the time differences with sub-second
resolution between the H$\alpha$ line emission observed with a
Multi-channel Subtractive Double Pass ({\it MSDP}) spectrograph on
the Large Coronagraph and Horizontal Telescope at Bia{\l}k\'ow
Observatory, Poland, and HXR emission recorded by the \emph{RHESSI}
spacecraft during several flares, greatly extending our earlier
analysis (Paper~I) to flares between 2003 and 2005.} {For 16
H$\alpha$ flaring kernels, observed in 12 solar flares, we made 72
measurements of time delays between local maxima of the
\emph{RHESSI} X-ray  and H$\alpha$ emissions. For most kernels,
there is an excellent correlation between time variations in the
H$\alpha$ line emission (at line centre and in the line wings) and
HXR (20--50~keV) flux, with the H$\alpha$ emission following
features in the HXR light curves generally by a short time lapse
$\Delta t = 1$--2~s, sometimes significantly longer (10--18~s). We
also found a strong spatial correlation. } {Owing to our larger
number of time measurements than in previous studies, the
distribution of $\Delta t$ values shows a much clearer pattern, with
many examples of short (1--2~s) delays of the H$\alpha$ emission,
but with some flares showing longer (10--18~s) delays. The former
are consistent with energy transfer along the flaring loop legs by
non-thermal electron beams, the latter to the passage of conduction
fronts.}

\maketitle

% Section 1
\section{Introduction}

In \cite{rad07}  (hereafter Paper I), we discussed high time
resolution observations of H$\alpha$ solar flares made at Bia{\l}kow
Observatory, Poland, and in hard X-rays in the 20--50 keV energy
range, as observed with the {\it Reuven Ramaty High-Energy Solar
Spectroscopic Imager} ({\it RHESSI}). Our aim was to establish time
correlations between features in the H$\alpha$ and {\it RHESSI}
light curves in the flare impulsive stage that will help us to
distinguish between mechanisms of energy transfer between the flare
energy release site in the corona and the source of H$\alpha$
emission in the chromosphere. These mechanisms are likely to be
either in the form of electron beams or conduction fronts from hot
plasmas.

Theoretical investigations \citep{Can87} indicate that intense
electron beams travelling from the energy release site to the
chromosphere would give rise to explosive evaporation in which
heated plasma pushes down on the chromosphere, leading to increases
in the H$\alpha$ line emission in the line centre and its red wing
after only $\sim 2$~s. Similar results were obtained by Heinzel
(1991), though he predicts dips before the main rise in emission,
both at the H$\alpha$ line centre and in the wings (H$\alpha \pm
1$~\AA). On the other hand, Smith and Lilliequist (1979) did
calculations for a HXR loop-top source that they had assumed to be
thermal rather than non-thermal, deducing that conduction fronts
moving at approximately the local ion sound speed ($\sim 200$~km
s$^{-1}$) would move towards the chromosphere, so enhancing the
H$\alpha$ emission. The travel time for such a conduction front with
a typical loop size of a few thousand km would be about 10~s or
more.

Observations in H$\alpha$ have been made over the flare impulsive
stage by \citet{kae83}, \citet{kur88}, \citet{tro00}, \citet{wan00},
and \citet{kas05}. These observations were made of single flares
either over the entire H$\alpha$ line profile or at specific
wavelengths (e.g. at H$\alpha - 1.0$~\AA\ in the case of
\citet{kur88}). Observations by \citet{han04} and \citet{gra90} at
three positions across the H$\alpha$ profile, including the line
centre, were made with time resolutions ranging from 0.033~s to
1.4~s. A comparison of the H$\alpha$ light curves with emission in
either hard X-rays or microwave radio emission in many of these
investigations leads to values of $\Delta t$ (delays in the
H$\alpha$ emission compared with either HXR or microwave radio
emission) of between a fraction of a second and several seconds. Our
own observations reported in Paper~I found that the H$\alpha$
emission is delayed with respect to HXR observed by {\it RHESSI} by
2--3~s for two flares, and up to 17~s for a third.

\begin{table*}
\begin{center}
\caption{Flares analysed in this paper.}
\end{center}
\vspace{-0.75cm}
\begin{tabular}{cccccccccc}
  \hline
  \hline
             &   Data      & H$\alpha$ Observations & Active  & Location & {\em GOES} & H$\alpha$ Cadence & H$\alpha$ & RHESSI & H$\alpha$ Kernel             \\
             &             & Start-End [UT]         & Region  &          &  Class     & [s]               & Telesc.   & Observ.&                              \\
   \hline
   1.        & 2003 Jul~16 & 15:57:45-16:06:05      & 10~410  & S10 E28  & C1.2       & 0.050             & LC        & yes    & {\bf K1}, {\bf K2}, {\bf K3} \\
%   \hline
   2.        & 2004 Apr~23 & 05:49:17-06:01:48      & 10~597  & S06 W83  & B9.1       & 0.075             & LC        & yes    & {\bf K4}                     \\
%   \hline
   3.        & 2004 Apr~23 & 09:28:50-09:41:19      & 10~597  & S07 W83  & C4.4       & 0.075             & LC        & yes    & {\bf K5}, {\bf K6}, K7       \\
%   \hline
   4.        & 2004 May~03 & 07:24:15-07:34:18      & 10~601  & S08 W54  & B2.5       & 0.060             & LC        & yes    & \textbf{K9}, K10                      \\
%   \hline
   5.        & 2004 May~21 & 05:44:08-05:50:48      & 10~618  & S10 E55  & C2.0       & 0.040             & LC        & yes    & {\bf K13}                    \\
%   \hline
   6.        & 2004 May~21 & 10:25:26-10:30:06      & 10~618  & S10 E55  & B7.0       & 0.040             & HT        & yes    & {\bf K14}                    \\
%   \hline
   7.        & 2005 Jan~17 & 08:00:59-08:11:59      & 10~720  & N13 W29  & X3.8       & 0.066             & LC        & part   & {\bf K15}, K16, K17, K18     \\
%   \hline
   8.        & 2005 Jul~12 & 07:53:10-08:10:20      & 10~786  & N09 W68  & C8.3       & 0.050             & HT        & yes    & {\bf K19}, {\bf K20}         \\
%   \hline
   9.       & 2005 Jul~12 & 10:00:44-10:09:04      & 10~786  & N09 W68  & C2.3       & 0.050             & HT        & part   & {\bf K21}                    \\
%   \hline
   10.       & 2005 Jul~12 & 13:02:11-13:10:30      & 10~786  & N09 W68  & M1.0       & 0.050             & HT        & part   & {\bf K25}, K26               \\
%   \hline
   11.       & 2005 Jul~13 & 08:15:04-08:26:09      & 10~786  & N11 W79  & C2.7       & 0.066             & HT        & yes    & K34, K35, {\bf K36}, K37     \\
%   \hline
   12.       & 2005 Jul~13 & 10:05:40-10:16:45      & 10~786  & N11 W79  & C1.6       & 0.066             & HT        & part   & {\bf K38}                    \\
   \hline
\end{tabular}

Note: Kernels indicated by bold type were used to measure $\Delta t$
(delay of H$\alpha$ emission relative to {\em RHESSI} 20--50~keV);
for the K9 kernel of the flare on 2004 May 3, measurements of
$\Delta t$ were only possible for energies $<20$~keV.

\end{table*}

For a more systematic study of the delay times, many more high time
resolution H$\alpha$ flare observations with simultaneous HXR
observations are needed. While three flares were discussed in Paper
I, in the present paper we include observations of 12 flares
containing one or more small bright regions (called here kernels) in
their H$\alpha$ emission, whose {\em GOES} X-ray importance ranges
from B to X. These observations, made at nine wavelengths across the
H$\alpha$ line profile and with high time resolution
(0.04--0.075~s), represent a considerable improvement over previous
investigations, most of which discuss single flares with more modest
time resolution and wavelength discrimination. They allow us to
compare, for the first time, the distribution of delay times of
recognizable features in the H$\alpha$ light curves with those in
hard X-ray light curves. From these data, we are able to state much
more definitively than hitherto the origin of the H$\alpha$ emission
features.

% Section 2
\section{Observational data and data analysis}

Our H$\alpha$ observations were taken with either the Large
Coronagraph (LC) or Horizontal Telescope (HT) using both the
Multi-channel Subtractive Double Pass (MSDP) spectrograph
\citep{mei91, rom94} and fast CCD cameras, all instruments and
telescopes being located at Bia{\l}kow Observatory, University of
Wroc{\l}aw, Poland. The CCD cameras form part of the Solar Eclipse
Coronal Imaging System \citep[SECIS:][]{phi00,rud04,rud10}, used
during solar eclipses. Our objective was to compare H$\alpha$ flare
observations with those of the X-ray emission made with {\it
RHESSI}. Owing to spacecraft night and SAA passages, {\it RHESSI}
data were available for only selected H$\alpha$ observations from
the full set of high-cadence observations with the {\it MSDP-SECIS}
instrument of solar flares or chromospheric brightenings over three
summer seasons, 2003--2005. This resulted in 12 flares having {\it
MSDP-SECIS} observations of H$\alpha$ kernels that displayed
brightness variations apparently similar to those in HXRs
(investigated in detail here and indicated in bold type in Table~1),
though some others were not. Details are given in Table~1, while
information about all our H$\alpha$ observations (including those
with no {\it RHESSI} data available) are given in Table~3. The
numbering scheme of the kernels follows the notation of this table.

%Fig. 1
\begin{figure}[!htp]
\includegraphics[width=8.7cm]{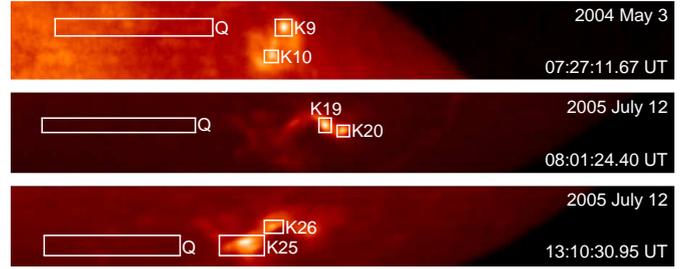}
\caption{H$\alpha$ line centre images of the solar flares on 2004
May 3 and 2005 July 12 observed with the {\it LC-MSDP-SECIS} or {\it
HT-MSDP-SECIS} systems at Bia{\l}k{\'o}w Observatory. The H$\alpha$
emission sources (H$\alpha$ kernels) are marked: K9, K10, K19, K20,
K25 and K26, while relevant reference quiet-chromosphere regions are
labelled Q. The field of view in these images is
325$\times$41~arcsec$^2$ (for top panel) and
942$\times$119~arcsec$^2$ (for middle and bottom panels),
respectively. Main characteristics of all the analysed flares are
given in Table 1.} \label{Fig01}
\end{figure}

The instrumentation has already been described in Paper~I so we give
only a brief summary here. The {\it MSDP} spectrograph has a
rectangular entrance window covering an area
325$\times$41~arcsec$^2$ on the Sun. A nine-channel prism-box
enables data to be collected in the range $\pm$1.6~\AA~ centred on the
H$\alpha$ line centre, resulting in either quasi-monochromatic
images in several wavelengths across the H$\alpha$ line profile
(line centre $\pm 1.2$~\AA\ with a band-width of 0.06~\AA\ for each
channel) or complete H$\alpha$ line profiles for all pixels in the
field of view. The images are recorded by one of the two CCD cameras
of the {\it SECIS} system. The other CCD camera is used to record
the precise time from the DCF77 long-wavelength transmitter (the
signal of which is generated by a caesium clock). The two {\it
SECIS} cameras have $512 \times 512$ pixel$^2$ CCD image sensors
(the photometric characteristics of the cameras were discussed by
\citet{rud10}) with an image scale $\sim 1$ arcsec per pixel. Up to
$10\,000$ images can be taken and stored using a specially adapted
computer with proprietary software.

We observed several H$\alpha$ flares having one or more kernels with
time cadences of between 0.04~s (25 images s$^{-1}$) and 0.075 s
($\sim$13 images s$^{-1}$), depending on the brightness of the
observed features. Corrections for small image displacements caused
by atmospheric seeing were made by a two-dimensional
cross-correlation of well-defined features (e.g. a sunspot visible
in far wings of the H$\alpha$ line), giving a positional accuracy of
$\sim 1$ pixel. Light curves of the emission of small rectangular
areas enclosing each kernel at 13 wavelengths (line centre $\pm
1.2$~\AA) across the H$\alpha$ profile were constructed (see
Fig.~\ref{Fig01}). A small but generally negligible amount of
non-flaring chromospheric emission was included in each area. To
compensate for brightness changes caused by seeing effects, the
flare kernel emission was normalized to the emission observed in a
neighbouring region of the quiet chromosphere. These areas are
indicated by ``Q" in Fig.~\ref{Fig01}.

%Fig. 2
\begin{figure*}[!htp]
\begin{center}
\includegraphics[width=17.5cm]{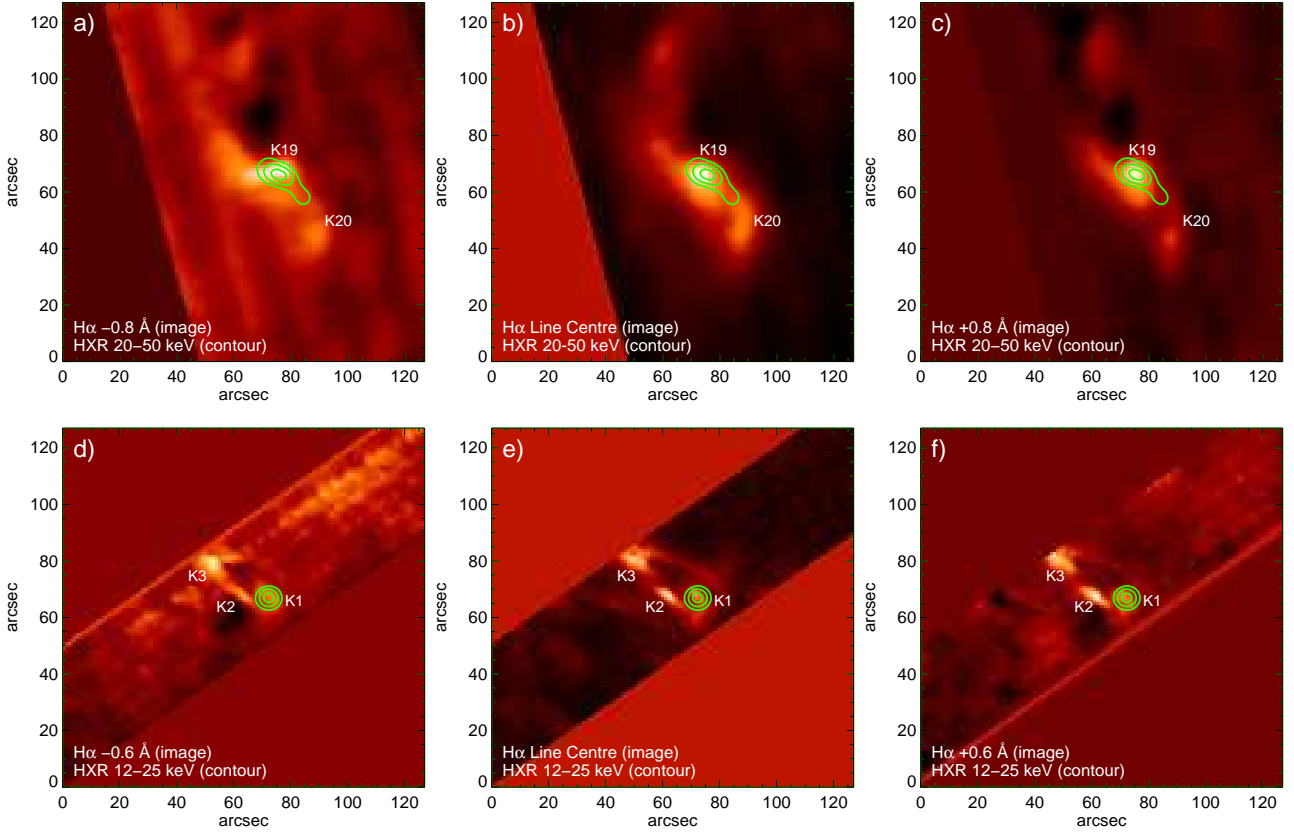}
\caption{H$\alpha$ and {\it RHESSI} HXR images (reconstructed using
the PIXON method, detectors: 3, 4, 5, 6, 8, 9) co-aligned as
described in the text. North is towards the top. The {\it RHESSI}
images are shown as contours (at 50$\%$, 70$\%$, and 90$\%$ of
maximum of signal). Panels: a,b,c: C8.3 flare - 2005 July 12,
08:00:52-08:00:56 UT ({\it RHESSI} 20-50 keV contours), 08:00:54 UT
H$\alpha$ image. Panels: d,e,f: C1.2 flare - 2003 July 16, 16:03:44
- 16:03:52 UT ({\it RHESSI} 12-25 keV contours), 16:03:46 UT
H$\alpha$ image. The dimensions of the areas shown are
128$\times$128 arcsec$^2$. } \label{Fig02}
\end{center}
\end{figure*}

Light curves of the flare X-ray emission, generally in the energy
range 20--50~keV, were obtained from {\it RHESSI} data. (Detectors
1, 3, 4, 5, 6, 7, 8 and 9 were used.) The time resolution of {\it
RHESSI} is normally determined by the spacecraft spin period (4~s),
which is rather too poor for detailed comparison of the flare
impulsive stage with the H$\alpha$ data. However, using a
demodulation procedure written by G. Hurford \citep{hur04}, the time
resolution was improved to 0.25~s. For weaker flares, the {\it
RHESSI} X-ray emission had poor statistical quality at this
demodulation level, thus a box-car smoothing to a 1-s time
resolution was applied.

%Fig. 3
\begin{figure}[!htp]
\includegraphics[width=8.7cm]{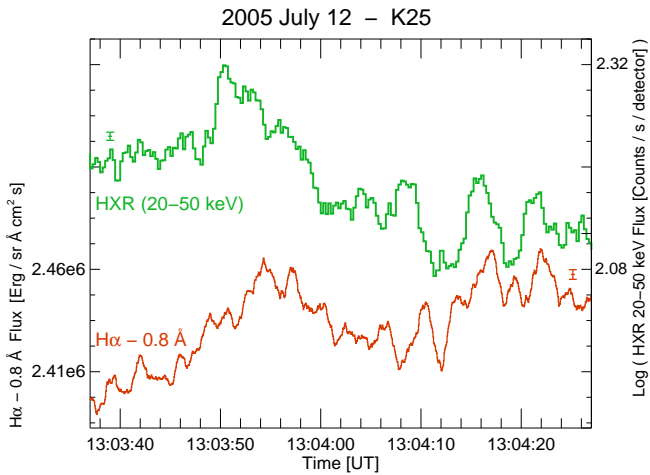}
\caption{Time variations in the {\it RHESSI} HXR (20--50 keV) solar
integrated flux (green histogram) and {\it HT-MSDP-SECIS} H$\alpha$
fluxes (red histogram) of the K25 flaring kernel recorded during the
M1.0 flare in NOAA 10\,786 active region on 2005 July~12. The
H$\alpha$ flux were taken $-0.8$~\AA\ from the line centre and are
on a linear scale. The {\it RHESSI} data are count rates (per second
per detector) on a logarithmic scale. The vertical scales indicate
the units of each. The integration times are 0.25~s for the
demodulated HXR data, and 0.05~s for H$\alpha$ data. The data were
smoothed using a 1-s box-car filter for HXR flux and 0.5-s box-car
filter for H$\alpha$ light curves. The error bars indicate the
standard deviations calculated for both light curves. Here and in
the next two figures the error bars in the X-ray light curves are
plotted on the left, while the error bars of the H$\alpha$ light
curves are plotted on the right.} \label{Fig03}
\end{figure}

The reality of variations in the H$\alpha$ emission was established
by whether they significantly exceeded the standard deviation in
random variations in the amount of emission in the non-flaring areas
(marked Q) in Fig.~\ref{Fig01}. The standard deviations $\sigma$ are
indicated by error bars in the H$\alpha$ light curves in
Figs.~\ref{Fig03}, \ref{Fig04}, and \ref{Fig05}. The uncertainties
in the {\em RHESSI} light curves are not simply given by those in
the count rates assuming Poissonian distributions because of the use
of the demodulation technique and (for weaker flares) subsequent
smoothing, so we estimated uncertainties in portions of the light
curve for each flare showing random variations observed during the
pre-flare phase. The combined uncertainties in the {\em RHESSI} and
H$\alpha$ emission were then used as a guide to identifications of
real features in the light curves. A feature was deemed real if it
was at least $3\sigma$ above neighbouring emission. Estimations of
time lags $\Delta t$ between the HXR and H$\alpha$ emission of
features were then made by eye. This procedure was found to be a far
more sensitive means of evaluating the time differences $\Delta t$
than using cross-correlation methods. Cross-correlation coefficients
$r$ were evaluated, however, as an indication of the degree of
similarity between the H$\alpha$ and {\em RHESSI} light curves.
These are indicated in Fig.~\ref{Fig04}.

The connection between the {\em RHESSI} and H$\alpha$ emission for
particular events from their relative timings was further confirmed
by using {\em RHESSI} imaging data for the particular time interval.
Data from {\em RHESSI} detectors 3, 4, 5, 6, 8, and 9 were used for
this, with the PIXON method of image re-construction (\cite{hur02}),
which is the method that constructs the simplest model of a {\em
RHESSI} image consistent with the data. The positions of the {\em
RHESSI} flare emission are obtained in standard solar coordinates,
but the precise positions of the H$\alpha$ emission require the
scale and rotation of the H$\alpha$ images to be determined. This
was done by comparing our H$\alpha$ images with data from the {\em
SOHO} MDI continuum channel, where sunspots common to both images
allow co-alignment to be established and therefore standard solar
coordinates of the H$\alpha$ flare emission with an estimated
accuracy of 2~arcsecs. Examples of co-aligned H$\alpha$ and {\em
RHESSI} images are shown for two flares in Fig.~\ref{Fig02}.

% Section 3
\section{Results}

For 12 flares, we analysed 16 H$\alpha$ kernels suitable for
measuring $\Delta t$ defined by the time difference between a
recognizable local maximum in the H$\alpha$ light curve (at line
centre or the blue or red wings) and the time of a corresponding
feature in the {\it RHESSI} light curve during the flare impulsive
stage. In total, 72 measurements of time delays $\Delta t$ were made.

We describe the evolution of three flares in particular. All these
flares have a wide range of {\em GOES} importance, and  our results
reflect the methods we used to evaluate the H$\alpha$ delay times
$\Delta t$. We then discuss the measured delay times for all the
analysed flares and the flare number distributions.

% Subsection 3.1
\subsection{M1 flare of 2005 July 12 (13:02-13:10 UT)}

The {\it RHESSI} light curve and corresponding H$\alpha$ light curve
of flaring kernel K25 in this M1 flare in NOAA active region
$10\,768$ are shown in Fig.~\ref{Fig03}. The H$\alpha$ image of this
kernel and K26 is shown in Fig.~\ref{Fig01} (bottom panel). The
H$\alpha$ data were taken in the blue wing ($-0.8$~\AA) of the line
and are on a linear scale. The integration time was 0.05~s. The {\it
RHESSI} HXR (20-50 keV) data are photon count rates on a logarithmic
scale. We used the \cite{hur04} demodulation procedure, but as the
resulting light curve was somewhat noisy, we applied smoothing with a 1-s
box-car filter. This is the light curve shown in
Fig.~\ref{Fig03}.

As can be seen in this figure, between 13:03:50 and 13:04:25 UT up
to five peaks in the {\it RHESSI} (20--50 keV) light curve can be
recognized, all with corresponding features in the H$\alpha$
$-0.8$~\AA\ light curve. The H$\alpha$ peaks have a significance of
several standard deviations (indicated in the figure). This allows
us to make five separate estimations of $\Delta t$, the time that
the H$\alpha$ peak is delayed with respect to the HXR peak, ranging
from 1 to 4~seconds.

% Subsection 3.2
\subsection{C8.3 flare of 2005 July 12 (07:53-08:10 UT)}

\begin{figure*}[!htp]
\begin{center}
\includegraphics[width=18.0cm]{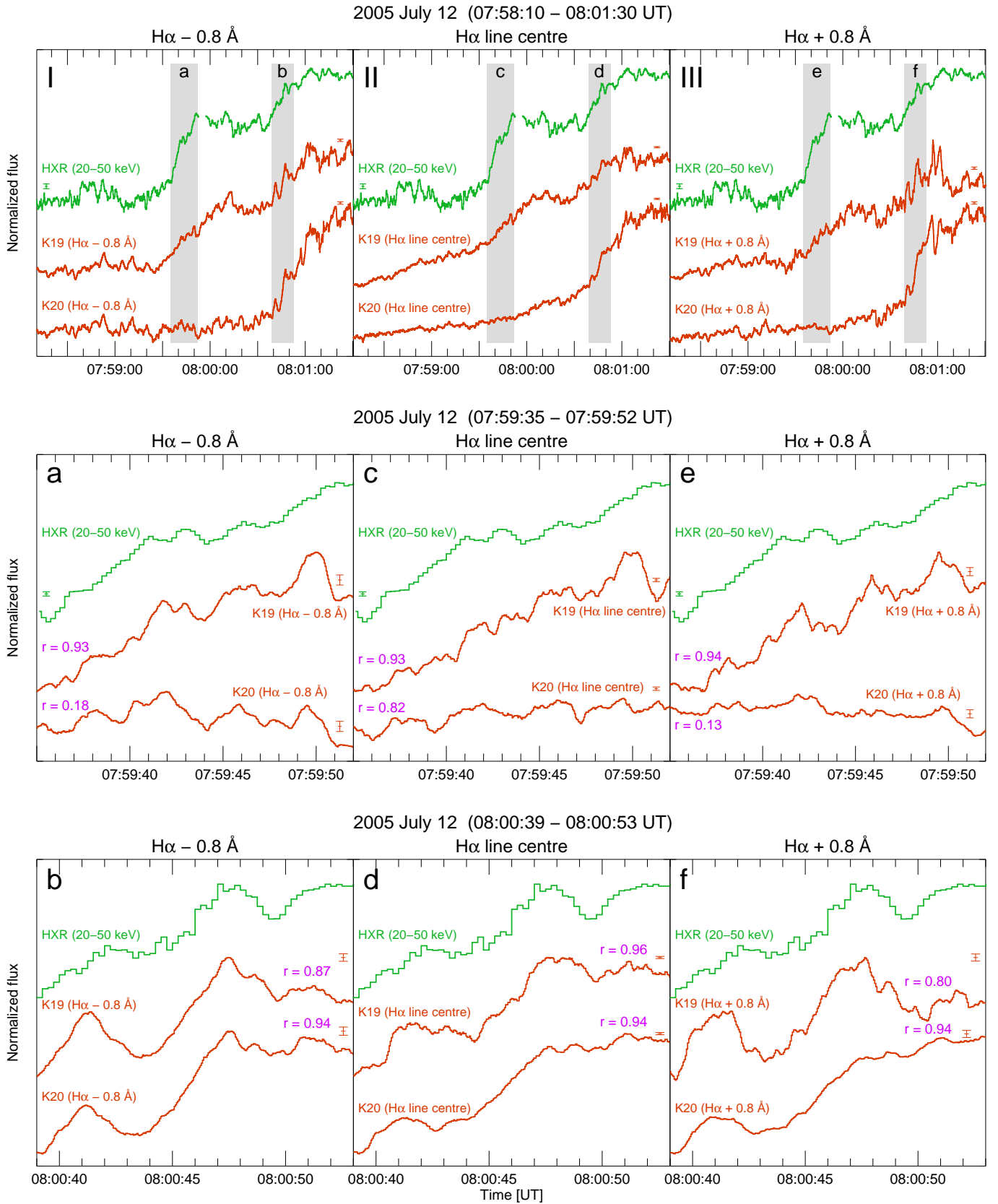}
\caption{The {\it RHESSI} HXR (20--50 keV) and {\it HT-MSDP-SECIS}
H$\alpha$ light curves of the K19 and K20 flaring kernels recorded
during the C8.3 on 2005 July~12. The H$\alpha$ light curves are
measured at line centre and $\pm 0.8$~\AA\ from the line centre and
plotted using a linear scale. The {\it RHESSI} data are count rates
(per second per detector) on a logarithmic scale. The integration
times are 0.25~s for the demodulated HXR data, and 0.05~s for
H$\alpha$ data. The data were smoothed using a 1-s box-car filter
for the HXR flux and 0.5-s box-car filter for the H$\alpha$ light
curves. The vertical grey strips in the upper panel (labelled a--f)
are magnified in the middle and lower panels of this figure. The
error bars indicate the standard deviations calculated for H$\alpha$
and HXR light curves, except panels b, d, and f where the
uncertainties in the HXR curves could not be easily established
because of the {\em RHESSI} A1 attenuator insertion at 07:59:56 UT.
The cross-correlation coefficients \emph{r} (between HXR and
H$\alpha$ fluxes) are shown in panels a--f. All light curves are
normalized to enhance the variations in intensity. The H$\alpha$ and
\emph{RHESSI} HXR energy flux ranges are defined in Table 2. }
\label{Fig04}
\end{center}
\end{figure*}

For this flare, two bright H$\alpha$ kernels (K19, K20) exhibited
significant time variations that are in some way similar to those in
the {\it RHESSI} 20--50~keV emission. The light curves of each are
shown in Fig.~\ref{Fig04}, while Fig.~\ref{Fig01} (middle panel)
shows our measurements for the two regions and the neighbouring emission of active region
$10\,786$ that was to produce the M1 flare discussed in Section 3.1
a few hours later.

In the top panel of Fig.~\ref{Fig04}, the H$\alpha$ fluxes are for
line centre and both wings, $\pm 0.8$~\AA\ from the line centre, on
a linear scale, while the {\it RHESSI} (20--50 keV) data are plotted
logarithmically; otherwise the figure is similar to
Fig.~\ref{Fig03}. The time intervals in the vertical grey strips
(upper panel, labelled $a$ to $f$) are shown magnified in the middle
and bottom panels. For this flare, the K19 and K20 kernels were
observed simultaneously. The brightness of K19 increases in all
parts of the H$\alpha$ profile simultaneously with the HXR (20-50
keV) flux increase, with several statistically significant ($>3
\sigma$) features present in all light curves. In contrast, the
brightness of the K20 kernel in the H$\alpha$ line blue wing
($-0.8$~\AA~from line centre) only starts to increase some
70~seconds after the K19 increase (see Figure~\ref{Fig04}, top-left
panel). However, the small variations in both the K19 and K20 light
curves between 07:59:35 UT and 07:59:52 UT in the H$\alpha$ blue
wing ($-0.8$~\AA) as well as in the red wing ($+0.8$~\AA) were
almost simultaneous (Fig.~\ref{Fig04}, left and right plots of the
middle panel). Both of these kernels display very similar variations
in the H$\alpha$ line wings and line centre. During the main
increase in the flare (between 08:00:39~UT and 08:00:53~UT), the
correlation between the X-ray and H$\alpha$ light curves measured at
the H$\alpha$ line centre as well as in both wings (at
$\pm~0.8$~\AA) was maintained (see Figure~\ref{Fig04}, panels $b$,
$d$, and $f$ in bottom panel). Almost all the small structures of
the HXR (20--50~keV) light curve have corresponding features in the
H$\alpha$ light curves, with very short time delays.

\begin{table*}
\begin{center}
\caption{The H$\alpha$ and \emph{RHESSI} HXR energy flux ranges in Figures 4 and 5.}
\end{center}
\vspace{-0.75cm}
\begin{tabular}{cccccccccc}
  \hline
  \hline
              Figure & H$\alpha$ Kernel & H$\alpha-0.8$~\AA\  & H$\alpha$ line centre                         &  H$\alpha+0.8$~\AA\ & \emph{RHESSI} 20--50 keV  \\
                     &                  & [erg sr$^{-1}$ \AA$^{-1}$ cm$^{-2}$ s$^{-1}$] & [erg sr$^{-1}$ \AA$^{-1}$ cm$^{-2}$ s$^{-1}$] & [erg sr$^{-1}$ \AA$^{-1}$ cm$^{-2}$ s$^{-1}$] & [counts/s/detector] \\
   \hline
   4 - upper row & K19 & $1.67 \cdot 10^6-2.22 \cdot 10^6$ & $1.11 \cdot 10^6-2.01 \cdot 10^6$ & $1.69 \cdot 10^6-2.37 \cdot 10^6$ & 31-2114  \\
     & K20 & $1.76 \cdot 10^6-2.20 \cdot 10^6$ & $0.98 \cdot 10^6-2.03 \cdot 10^6$ & $1.52 \cdot 10^6-2.08 \cdot 10^6$ &          \\
   4 - middle row  & K19 & $1.77 \cdot 10^6-1.89 \cdot 10^6$ & $1.38 \cdot 10^6-1.58 \cdot 10^6$ & $1.80 \cdot 10^6-1.97 \cdot 10^6$ & 83-842   \\
     & K20 & $1.78 \cdot 10^6-1.83 \cdot 10^6$ & $1.12 \cdot 10^6-1.18 \cdot 10^6$ & $1.56 \cdot 10^6-1.60 \cdot 10^6$ &          \\
   4 - lower row  & K19 & $1.96 \cdot 10^6-2.13 \cdot 10^6$ & $1.74 \cdot 10^6-1.93 \cdot 10^6$ & $2.02 \cdot 10^6-2.28 \cdot 10^6$ & 665-1328 \\
     & K20 & $1.83 \cdot 10^6-1.99 \cdot 10^6$ & $1.40 \cdot 10^6-1.70 \cdot 10^6$ & $1.67 \cdot 10^6-1.94 \cdot 10^6$ &          \\
   \hline
                    &                  & H$\alpha-0.6$~\AA\  & H$\alpha$ line centre                          &  H$\alpha+0.6$~\AA\ & \emph{RHESSI} 10--20 keV   \\
   \hline
   5 & K9  & $1.44.10^6-1.67 \cdot 10^6$       & $0.92 \cdot 10^6-1.1 \cdot 10^6$  & $1.36 \cdot 10^6-1.55 \cdot 10^6$ & 40-160   \\
   \hline
\end{tabular}\\
\end{table*}

% Subsection 3.3
\subsection{B2.5 flare of 2004 May 3 (07:24-07:34 UT)}

\begin{figure*}[!htp]
%\begin{center}
\includegraphics[width=17.0cm]{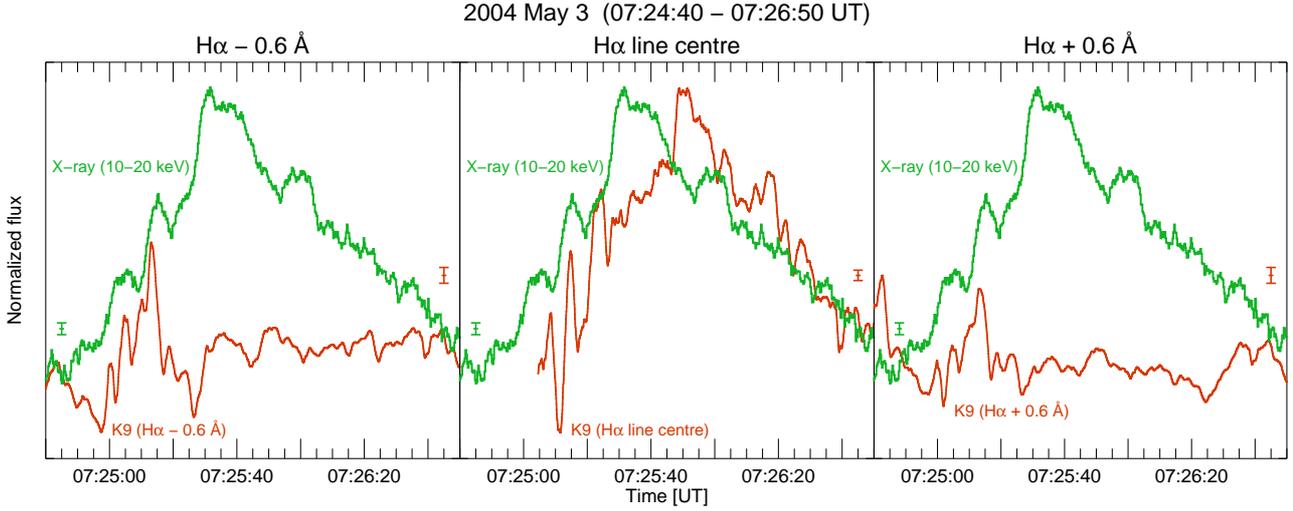}
\caption{The {\it RHESSI} (10-20 keV) and {\it LC-MSDP-SECIS}
H$\alpha$ light curves of the K9 flaring kernel during the B2.5
flare on 2004 May~3. The H$\alpha$ light curves are shown for line
centre and $\pm 0.6$~\AA\ from the line centre. The data in the
upper panel were smoothed using a 4-s box-car filter for the {\em
RHESSI} light curve and a 1-s box-car filter for the H$\alpha$ light
curves. The {\em RHESSI} light curves in the bottom panel were
smoothed using a 0.5-s box-car filter (thin curves) and a 4-s
box-car filter (thick curves), while the H$\alpha$ light curves
shown are unsmoothed (thin curves) and smoothed using a 1-s box-car
filter (thick curves). The vertical grey strips (labelled a--c) are
shown magnified in the bottom panel. The error bars indicate the
standard deviations calculated for H$\alpha$ and X-ray light curves.
The H$\alpha$ and \emph{RHESSI} HXR energy flux ranges are defined
in Table 2. } \label{Fig05}
%\end{center}
\end{figure*}

The correlation of the time variations in the HXR integrated flux
and H$\alpha$ light curves for separate flaring kernels are evident
even for relatively small flaring events, as in this B2.5 {\em
GOES}-class solar flare observed on 2004 May 3 (NOAA 10\,601). The
flare had two clearly separated H$\alpha$ flaring kernels (K9 and K10,
Fig.~\ref{Fig01}, top panel). However, unlike the C8.3 flare of 2005
July 12, the time variations in the H$\alpha$ emission of the K9
kernel, in various parts of the line, were different
(Fig.~\ref{Fig05}: H$\alpha$ light curves at line centre and in both
wings at $\pm~0.6$~\AA).

During the impulsive stage of the flare (up to 07:25:20~UT), the
H$\alpha$ light curves taken at the line centre and in both the line
wings display time variations similar to those of the X-ray
(10--20~keV) flux (the increase of HXR emission above 20~keV was too
weak to be recorded by {\em RHESSI}). Unexpectedly, after
07:25:20~UT the intensities of the H$\alpha$ light curves taken in
both line wings dropped suddenly, roughly to the pre-flare level,
and did not reveal any similarities to the X-ray light curve. During
the maximum and late phases of the flare, H$\alpha$ line-centre
emission was recorded only, showing a maximum $\sim$ 20~s after
maximum of the X-ray \mbox{(10--20~keV)} emission. This implies that
there is a rather slow transport of energy to the chromosphere that
is more consistent with the travel time of a conduction front moving
at the ion sound speed than electron beam travel times.

%Fig. 4

%Fig. 5

% Subsection 3.4
\subsection{Time delays of H$\alpha$ features in flares}

We selected 12 solar flares observed in {\it MSDP-SECIS} H$\alpha$
and {\it RHESSI} HXR (20--50~keV) ranges and investigated 16
kernels, suitable for measurements of the time delays $\Delta t$
between HXR and H$\alpha$ local maxima in the light curves. We were
able to make 72 measurements of the time delays $\Delta t$
consisting of 26 measurements for H$\alpha$ light curves taken in
the line centre, 24 measurements for light curves taken in blue
wing, and 22 measurements for light curves taken in the red wing.

\begin{figure}[h!]
\includegraphics[width=8.7cm]{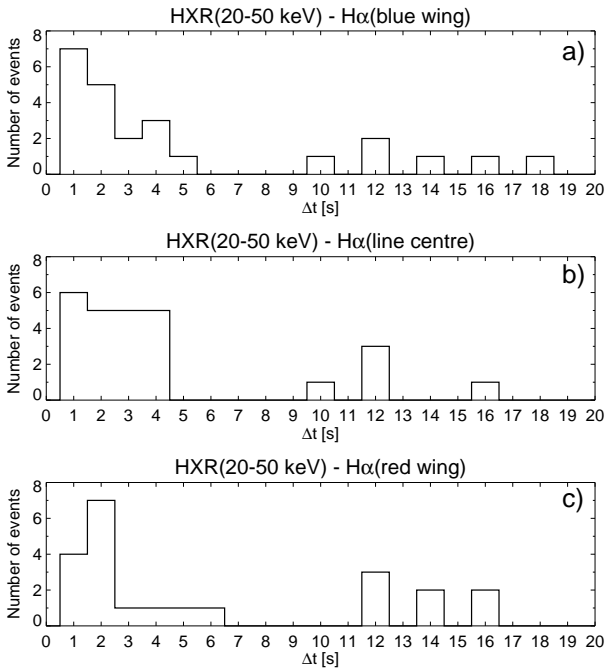}
\caption{Distribution of all 72 measurements of the time delays $\Delta$t
between {\it RHESSI} HXR (20-50 keV) and {\it MSDP-SECIS} H$\alpha$
localised maxima measured at the line centre  a), blue wing  b), and red
wing  c), respectively. } \label{Fig06}
\end{figure}

The histograms of the time delays $\Delta t$ between HXR
\mbox{(20--50~keV)} and H$\alpha$ local maxima of emission are shown
in Fig.~\ref{Fig06}. The three panels of this figure show relevant
time delays between local peaks of: HXR emission and emission
measured in the blue wing of the H$\alpha$ line (panel $a$), HXR
emission and H$\alpha$ line centre emission (panel $b$), and HXR
emission and emission measured in the red wing of the H$\alpha$ line
(panel $c$). All histograms show two clearly distinct groups of time
delays: many short delays \mbox{$\Delta t \sim 1$--6~s}, and a
smaller number of features with longer delays, \mbox{$\Delta t \sim
10$--18~s}.

%Fig. 6

For all flares with short time delays between HXR and H$\alpha$
($\Delta t <6$ seconds), the cross-correlation coefficients were
relatively high, from 0.7 up to 0.9. Over some short time periods
for certain flares, such as for the K19 kernel during the impulsive
stage of the 2005 July~12 ($\sim$ 08:00 UT) flare, the
cross-correlation coefficients were as high as 0.96
(Fig.~\ref{Fig04} - panel $d$).

% Section 4
\section{Discussion and conclusions}

The histograms of the time-delays show two clearly distinct groups
of time delays: many flares have short-time delays $\Delta t \sim
1$--6~s, and a few have $\Delta t \sim 10$--18~s. The short time
delays are consistent with the HXR and H$\alpha$ source regions
being closely separated, with energy transfer by non-thermal
electrons from the primary energy source to the chromosphere. The
time delays are comparable to those in the model calculations of
Heinzel (1991) and Kasparova et al. (2009); however, we did not
detect any reduction in H$\alpha$ emission immediately after the HXR
emission as Heinzel (1991) predicts. The longer lasting time delays
(10--18~s) may be explained by a slower energy transfer mechanism,
perhaps a conduction front moving with the ion sound speed ($\sim
200$~km~s$^{-1}$) down the flaring loop legs, as discussed in
Paper~I and by Trottet et al. (2000).

The H$\alpha$ light curves during the C8.3 solar flare on 2005
July~12 are particularly interesting. Although a brightening in the
K20 kernel, which is conspicuous in the H$\alpha$ blue (0.8~\AA) wing,
started about 70~s after an increase in the K19 light curve, smaller
variations in the K19 and K20 light curves measured in both the blue
and red H$\alpha$ wings (line centre $\pm 0.8$~\AA) were similar and
almost simultaneous. In addition, the {\em RHESSI} 20--50~keV
emission was highly correlated with the H$\alpha$ emission from at
least the K19 flaring kernel over the period 08:00:39--08:00:53~UT,
having a cross-correlation coefficient of 0.96 (Fig.~\ref{Fig04} -
panel d). This is consistent with the K19 and K20 kernels being
excited simultaneously by non-thermal electron beams travelling down
the flaring loop legs if K19 and K20 are the footpoints of the
flaring loop, perhaps asymmetrically since the H$\alpha$ response of
the K19 kernel was greater than that of K20. The delayed response of
the K20 kernel to the X-ray emission might have been due to
chromospheric evaporation at this location (e.g. \citet{fal09}),
though in the absence of either X-ray or ultraviolet line profile
data we are unable to confirm this.

The time variations of the H$\alpha$ and X-ray (10--20~keV) light
curves recorded during the B2.5 solar flare on 2004 May~3 may be
interpreted as energy transfer by electron beams before 07:25:20~UT
when the time delays of the H$\alpha$ variations are very short
($\Delta t < 2$ s) relative to the 10--20~keV X-ray emission.
These electron beams apparently reach deep into the chromosphere
where the H$\alpha$ line wings are formed. However, after 07:25:20
UT, the H$\alpha$ line centre light curve appears to be delayed by
$\sim 20$~s relative to the X-ray, suggesting a more gradual
energy transfer, perhaps by conduction which reaches only to the
upper chromosphere where the H$\alpha$ line centre is formed. We
note that similar impulsive variations in the soft X-ray emission (at
even lower energies, 0.6--3~keV) were recorded by \citet{hud94}.

The H$\alpha$ and {\it RHESSI} imaging data illustrates yet more
clearly the connection between the chromospheric and X-ray emission;
in particular, the images show the very close correspondence between
the hard X-ray and H$\alpha$ emission for flares that have very
small values of $\Delta t$.

In summary, the results presented here illustrate how hard X-ray and
H$\alpha$ observations with high time resolution offer much insight
into the mechanism whereby energy is transferred from the energy
release site in the corona to the chromosphere where the H$\alpha$
line is formed. The analysis reported here should provide impetus
for future observations with fast-frame CCD camera systems on solar
telescopes with spectral capabilities such as the system we have been
using at Bia{\l}k\'ow Observatory, and the ROSA imaging system
currently installed at the National Solar Observatory at Sacramento
Peak \citep{jes10}.

%\onltab{3}{
\begin{table*}
\begin{center}
\caption{The full list of high cadence {\it LC-MSDP-SECIS} and {\it
HT-MSDP-SECIS} H$\alpha$ observations made in 2003--2005 at the
Bia{\l}k{\'o}w Observatory (Wroc{\l}aw University, Poland). The
table includes: date and time of H$\alpha$ observations, NOAA active
region number, location on solar disk, {\em GOES}-class, cadence of
H$\alpha$ observations, the H$\alpha$ telescope used, availability
of {\it RHESSI} observations, and label for the H$\alpha$ emission
sources (H$\alpha$ kernels). The data sets used for analysis in this
paper are marked by bold types.}
\end{center}
\vspace{-0.75cm}
\begin{tabular}{cccccccccc}
  \hline
             &   Data            & H$\alpha$ Observations   & Active        & Location         & {\em GOES} & H$\alpha$ Cadence & H$\alpha$        & {\em RHESSI}    & H$\alpha$ Kernels      \\
             &                   & Start-End [UT]           & Region        &                  &  Class     & [s]               & Telesc.          & Observ.   &                        \\
   \hline

        1.   &      2003 Jul~16  &      07:00:41-07:09:01   &      10~410   &      S12 E30     &      B4.5  &      0.050        &      LC          &     no    &       -                \\
        2.   &      2003 Jul~16  &      11:05:57-11:14:16   &      10~410   &      S12 E30     &      B4.0  &      0.050        &      LC          &     yes   &       -                \\
   {\bf 3.}  & {\bf 2003 Jul~16} & {\bf 15:57:45-16:06:05}  & {\bf 10~410}  & {\bf S10 E28}    & {\bf C1.2} & {\bf 0.050}       & {\bf LC}         & {\bf yes} & {\bf K1, K2, K3}       \\
        4.   &      2003 Jul~17  &      10:11:57-10:20:17   &      10~410   &      S07 E22     &      B9.5  &      0.050        &      LC          &     yes   &       -                \\
        5.   &      2003 Jul~24  &      13:00:21-13:08:41   &      10~410   &      S13 W80     &      B8.5  &      0.050        &      LC          &     yes   &       -                \\
        6.   &      2003 Aug~21  &      07:52:44-07:59:51   &      10~436   &      N08 E23     &      B9.7  &      0.050        &      LC          &     no    &       -                \\
        7.   &      2003 Aug~25  &      15:18:08-15:22:45   &      10~436   &      N08 W36     &      B8.5  &      0.050        &      LC          &     yes   &       -                \\
        8.   &      2003 Aug~26  &      10:16:46-10:24:06   &      10~442   &      S14 E26     &      B7.3  &      0.050        &      LC          &     yes   &       -                \\
        9.   &      2003 Sep~03  &      06:00:44-06:09:04   &      10~448   &      N19 W47     &        -   &      0.050        &      LC          &     yes   &       -                \\
       10.   &      2004 Apr~15  &      09:49:52-10:02:22   &      10~591   &      S14 W34     &      B3.3  &      0.075        &      LC          &     no    &       -                \\
       11.   &      2004 Apr~22  &      10:55:25-11:03:45   &      10~597   &      S05 W77     &        -   &      0.050        &      LC          &     yes   &       -                \\
       12.   &      2004 Apr~22  &      15:10:18-15:22:48   &      10~597   &      S05 W77     &        -   &      0.075        &      LC          &     yes   &       -                \\
   {\bf 13.} & {\bf 2004 Apr~23} & {\bf 05:49:17-06:01:48}  & {\bf 10~597}  & {\bf S06 W83}    & {\bf B9.1} & {\bf 0.075}       & {\bf LC}         & {\bf yes} & {\bf K4}               \\
   {\bf 14.} & {\bf 2004 Apr~23} & {\bf 09:28:50-09:41:19}  & {\bf 10~597}  & {\bf S07 W83}    & {\bf C4.4} & {\bf 0.075}       & {\bf LC}         & {\bf yes} & {\bf K5, K6, K7}       \\
        15.  &      2004 Apr~23  &      11:49:38-11:56:18   &      10~597   &      S07 W84     &      M1.5  &      0.040        &      LC          &      part &  K8                    \\
   {\bf 16.} & {\bf 2004 May~03} & {\bf 07:24:15-07:34:18}  & {\bf 10~601}  & {\bf S08 W54}    & {\bf B2.5} & {\bf 0.060}       & {\bf LC}         & {\bf yes} & {\bf K9, K10}          \\
        17.  &      2004 May~05  &      08:38:30-08:46:50   &      10~605   &      S10 W10     &        -   &      0.050        &      HT          &      no   &  K11                   \\
        18.  &      2004 May~05  &      11:44:35-11:51:15   &      10~605   &      S10 W10     &      B5.4  &      0.040        &      HT          &      no   &  K12                   \\
        19.  &      2004 May~20  &      10:15:17-10:21:57   &      10~618   &      S08 E74     &      B2.8  &      0.040        &      HT          &      no   &       -                \\
        20.  &      2004 May~20  &      12:56:18-13:02:58   &      10~618   &      S10 E69     &      B3.5  &      0.040        &      LC          &      yes  &       -                \\
        21.  &      2004 May~20  &      16:22:55-16:29:35   &      10~618   &      S10 E69     &        -   &      0.040        &      LC          &      no   &       -                \\
   {\bf 22.} & {\bf 2004 May~21} & {\bf 05:44:08-05:50:48}  & {\bf 10~618}  & {\bf S10 E55}    & {\bf C2.0} & {\bf 0.040}       & {\bf LC}         & {\bf yes} &    {\bf K13}           \\
   {\bf 23.} & {\bf 2004 May~21} & {\bf 10:25:26-10:30:06}  & {\bf 10~618}  & {\bf S10 E55}    & {\bf B7.0} & {\bf 0.040}       & {\bf HT}         & {\bf yes} &    {\bf K14}           \\
        24.  &      2004 Sep~10  &      07:12:10-07:23:15   &      10~672   &      N05 E29     &      C1.7  &      0.066        &      LC          &     yes   &       -                \\
        25.  &      2004 Sep~10  &      07:24:17-07:35:22   &      10~672   &      N05 E29     &      C1.7  &      0.066        &      LC          &     part  &       -                \\
        26.  &      2004 Sep~10  &      12:56:23-13:07:28   &      10~672   &      N05 E29     &      C2.1  &      0.066        &      LC          &     part  &       -                \\
   {\bf 27.} & {\bf 2005 Jan~17} & {\bf 08:00:59-08:11:59}  & {\bf 10~720}  & {\bf N13 W29}    & {\bf X3.8} & {\bf 0.066}       & {\bf LC}         & {\bf part}&{\bf K15, K16, K17, K18}\\
        28.  &      2005 Jan~17  &      08:13:27-08:24:26   &      10~720   &      N13 W29     &      X3.8  &      0.066        &      LC          &     no    &       -                \\
        29.  &      2005 Jan~17  &      11:41:20-11:52:19   &      10~720   &      N13 W29     &      X3.8  &      0.066        &      HT          &     yes   &       -                \\
        30.  &      2005 Apr~01  &      14:41:38-14:52:35   &      10~745   &      N12 W94     &        -   &      0.066        &      LC          &     part  &       -                \\
        31.  &      2005 Apr~02  &      08:26:18-08:37:17   &      10~747   &      S09 E32     &        -   &      0.066        &      LC          &     no    &       -                \\
        32.  &      2005 Apr~02  &      08:37:50-08:48:50   &      10~747   &      S09 E32     &        -   &      0.066        &      LC          &     no    &       -                \\
        33.  &      2005 Jun~29  &      09:06:39-09:14:59   &      10~781   &      N15 E51     &        -   &      0.050        &      LC          &     no    &       -                \\
   {\bf 34.} & {\bf 2005 Jul~12} & {\bf 07:53:10-08:01:30}  & {\bf 10~786}  & {\bf N09 W68}    & {\bf C8.3} & {\bf 0.050}       & {\bf HT}         & {\bf yes} & {\bf K19, K20}         \\
   {\bf 35.} & {\bf 2005 Jul~12} & {\bf 08:02:00-08:10:20}  & {\bf 10~786}  & {\bf N09 W68}    & {\bf C8.3} & {\bf 0.050}       & {\bf HT}         & {\bf yes} & {\bf K19, K20}         \\
   {\bf 36.} & {\bf 2005 Jul~12} & {\bf 10:00:44-10:09:04}  & {\bf 10~786}  & {\bf N09 W68}    & {\bf C2.3} & {\bf 0.050}       & {\bf HT}         & {\bf part}& {\bf K21}              \\
        37.  &      2005 Jul~12  &      12:10:15-12:18:34   &      10~786   &      N09 W68     &      C1.5  &      0.050        &      HT          &      no   & K22, K23, K24          \\
   {\bf 38.} & {\bf 2005 Jul~12} & {\bf 13:02:11-13:10:30}  & {\bf 10~786}  & {\bf N09 W68}    & {\bf M1.0} & {\bf 0.050}       & {\bf HT}         & {\bf part}& {\bf K25, K26}         \\
        39.  &      2005 Jul~12  &      15:27:57-15:39:01   &      10~786   &      N09 W68     &      C2.3  &      0.066        &      HT          &      no   & K27, K28, K29, K30, K31\\
        40.  &      2005 Jul~12  &      15:39:34-15:50:39   &      10~786   &      N09 W68     &      M1.5  &      0.066        &      HT          &      no   & K32,K33                \\
   {\bf 41.} & {\bf 2005 Jul~13} & {\bf 08:15:04-08:26:09}  & {\bf 10~786}  &      N11 W79     & {\bf C2.7} & {\bf 0.066}       & {\bf HT}         & {\bf yes} &{\bf K34, K35, K36, K37}\\
   {\bf 42.} & {\bf 2005 Jul~13} & {\bf 10:05:40-10:16:45}  & {\bf 10~786}  &      N11 W79     & {\bf C1.6} & {\bf 0.066}       & {\bf HT}         & {\bf part}& {\bf K38}              \\
        43.  &      2005 Jul~13  &      12:04:40-12:15:45   &      10~786   &      N11 W79     &      M3.2  &      0.066        &      HT          &      no   & K39, K40, K41          \\
        44.  &      2005 Jul~13  &      12:16:11-12:27:16   &      10~786   &      N11 W79     &      M3.2  &      0.066        &      HT          &      no   & K39, K40, K41          \\
        45.  &      2005 Jul~14  &      16:05:14-16:16:19   &      10~786   &      N11 W91     &      C4.5  &      0.066        &      HT          &      part &       -                \\
        46.  &      2005 Aug~26  &      11:42:21-11:50:41   &      10~803   &      N12 E53     &      C2.1  &      0.050        &      LC          &      no   & K42                    \\
        47.  &      2005 Aug~26  &      11:50:57-11:59:17   &      10~803   &      N12 E53     &      C2.1  &      0.050        &      LC          &      no   & K42                    \\
   \hline
\end{tabular}
\end{table*}
%}

% Section 5
%\section{Acknowledgements}

\end{document}